\documentclass[12pt]{article}
\usepackage{amsmath} 
\usepackage{amssymb}
\usepackage{hhline}
\usepackage[scale={.75,.75}]{geometry}
\usepackage{url}
\usepackage{lscape}
\usepackage{caption}
\usepackage[numbers,sort&compress]{natbib}
\usepackage{epsfig}
\usepackage{multirow} 
\usepackage{color}
\usepackage{verbatim}
\usepackage{nicefrac}
\usepackage{upgreek}
\usepackage{bbm}
\usepackage{setspace}
\usepackage{pstricks}
\usepackage{psfrag}
\usepackage{color}

\begin{document}
\date{}

\title{\vspace{-1.5cm} 
\begin{flushright}
\vspace{-0.4cm}
\end{flushright}
{ \bf   Gravitational Wave Imprint of New Symmetry Breaking }
}

\author{
\vspace{0.2cm}
Wei Chao$^a$\footnote{Email address: chaowei@bnu.edu.cn}, ~~
Wen-Feng Cui$^b$\footnote{Email address: cuiwenfeng@itp.ac.cn}, ~~
Huai-Ke Guo$^b$\footnote{Email address: ghk@itp.ac.cn},~~~
Jing Shu$^{b,c,d}$\footnote{Email address: jshu@itp.ac.cn}
\\
\footnotesize{$^a$Center for Advanced Quantum Studies, Department of Physics, Beijing Normal University, }\\
\footnotesize{ Beijing, 100875, China}\\
\footnotesize{$^b$CAS Key Laboratory of Theoretical Physics, Institute of Theoretical Physics,
}\\
\footnotesize{Chinese Academy of Sciences, Beijing 100190, China}
\\
\footnotesize{$^{c}$
CAS Center for Excellence in Particle Physics, Beijing 100049, China
}
\\
\footnotesize{$^{d}$
School of Physical Sciences, University of Chinese Academy of Sciences, Beijing 100190, P. R. China
}
}
\maketitle

\begin{abstract}
\noindent   
It is believed that there are extra fundamental gauge symmetries beyond these described by the Standard Model of particle physics.
The scale of these new gauge symmetries  are usually too high to be reachable by particle colliders. 
Considering that the phase transition (PT) relating to the spontaneous breaking of new gauge symmetries to the electroweak symmetry might be strongly first order, we propose in this paper taking the stochastic gravitational waves (GW) arising from this phase transition as an indirect  way of detecting these new fundamental gauge symmetries. 
As an illustration, we explore the possibility of detecting the stochastic GW generated from the PT of $\mathbf{ B-L}$ in the space-based interferometer detectors. 
Out study shows that the GW energy spectrum is reachable by the  LISA, BBO, Taiji and DECIGO experiments only for the case where the spontaneous breaking of $\mathbf{B-L}$ is triggered by at least two electroweak singlet scalars.

\end{abstract}

\section{Introduction}\label{sec:introduction}

Although predictions of the Standard Model (SM) of particle physics remarkably agree with almost all experimental observations, we never stop exploring new fundamental gauge symmetries beyond these described by the SM, which are usually motivated by the neutrino masses, dark matter, baryon asymmetry of the universe and  the gauge couplings unification at a Grand Unified Theory (GUT).
Scales relevant to the spontaneous breaking of new symmetries  are usually too high to be accessible by colliders in a foreseeable future. 
How to probe them is  an open question.  

The observation of Gravitational Wave (GW) signal at the Laser Interferometer Gravitational Wave Observer (LIGO)~\cite{TheLIGOScientific:2016pea} has opened a new window to explore the universe and various mysteries of particle physics~\cite{Ivanov:2017dad,Beniwal:2017eik,Cai:2017cbj,Addazi:2017gpt,Tsumura:2017knk,Kang:2017mkl,Chao:2017oux,Bian:2017wfv,Huang:2017rzf,Addazi:2017oge,Marzola:2017jzl,Chiang:2017zbz,Chao:2017vrq}. 
There are usually two sources of GW~\cite{Cai:2017cbj}: (1) cosmological origin, such as inflation and phase transition (PT);  (2) relativistic astrophysical origin ( Binary systems etc.). 
If phase transitions related to the spontaneous breaking of the new gauge symmetries are strongly first order, bubbles of broken phase may nucleate in the background of symmetric phase when the universe cools down to  the bubble nucleation temperature. 
Bubbles expand, collide, merge and finally fill the whole universe to finish the PT, and stochastic GW signals can be generated via the bubble collisions, sound waves after the bubble collision and turbulent motion of bulk fluid~\cite{Caprini:2015zlo}. 
In this paper we propose taking GW as an indirect way of exploring new gauge symmetries,  supposing the PT of new gauge symmetry breaking is strongly first order.  

Considering the complexity of the non-Abelian gauge group extended models, we study GWs generated from PTs of Abelian gauge group extended models  in this paper. 
There are many possible $U(1)$ extensions of the SM~\cite{Langacker:2008yv}, of which gauged $\mathbf{B-L}$~\cite{Mohapatra:1980qe,Marshak:1979fm,Wetterich:1981bx}, $\mathbf{B}$, $\mathbf{L}$~\cite{FileviezPerez:2010gw,Dulaney:2010dj,Chao:2010mp}, $\mathbf{B+L}$~\cite{Chao:2015nsm,Chao:2016avy}, $\mathbf{L_i-L_j}$~\cite{He:1991qd} (Here $\mathbf{B}$ and $\mathbf{L}$ are the baryon number and lepton number, respectively) have received great attentions. 
Since $U(1)_{\mathbf{B-L}}$ only need minimal extensions to the SM for anomalies cancellation, it is believed to be the most natural one according to Occam's Razor\footnote{Notice that the $U(1)_\mathbf{R}$~\cite{Chao:2017rwv}, the gauge symmetry for right-handed fermions, shares the same merit as $U(1)_\mathbf{B-L}$ on anomalies cancellation, but this model is severely constrained by the $Z-Z^\prime$ mixing.}.
We investigate conditions for the bubble nucleation during the PT of $U(1)_{\mathbf{B-L}}$, then calculate the energy spectrum of GWs generated from this process. 
Notice that the higher  the energy scale of PT is,  the larger peak frequency of GW energy spectrum  it has~\cite{Dev:2016feu}.
If $U(1)$ is broken at the TeV scale, its GW can be detected at  the space-based laser interferometer detectors such as  the Laser Interferometer Space Antenna(LISA), Big Bang Observer (BBO), Taiji and Tianqin projects. 
Alternatively if $U(1)$ is broken at a scale approaching to the GUT, its GW is sensitive to the ground-based Laser interferometer such as  aLIGO.
Our results show that  it is difficult to get large enough GW energy spectrum reachable by the space-based Laser interferometer if the $\mathbf{B-L}$ is broken by only one electroweak scalar singlet. 
Alternatively if $\mathbf{B-L}$ is broken by at least two electroweak scalar singlets,  its GW energy spectrum is detectable by the LISA detector, ALIA, DECIGO, BBO and Ultimate-DECIGO.
For GWs from the spontaneous breaking of non-Abelian symmetries, we refer the reader to Ref.~\cite{Huang:2017laj} for the case of 3-3-1 model~\cite{Pisano:1991ee,Frampton:1992wt}.

The remaining of the paper is organized as follows: In section~\ref{sec:model} we give a brief introduction to the Abelian gauge group extensions to the SM and describe the $U(1)_{\mathbf{B-L}}$ model in detail. 
Section~\ref{sec:GWs} is focused on the GW signals from the PT of $U(1)_\mathbf{B-L}$.
The last part is concluding remarks.

\section{Abelian gauge group extensions to the SM}\label{sec:model}

Many $U(1)$ extensions to the SM have been proposed in recent years, often with the motivation of resolving problems in cosmology and astrophysics.  
There are two ways to construct a gauged $U(1)$ symmetry:  top-down approach and bottom up approach.
A typical example of top-down approach is $U(1)$ from the $E_6$ GUT~\cite{King:2005jy}. 
At the GUT scale, $E_6$ can be broken directly into $SU(3)_C \times SU(2)_L \times U(1)_Y\times U(1)_\psi\times U(1)_\chi$ via the Hosotani mechanism~\cite{Hosotani:1983xw}.
Some phenomena inspired  U(1), such as $\mathbf{L_i-L_j}$, general $U(1)$~\cite{Appelquist:2002mw}, $U(1)_N$~\cite{Chao:2012sz,Chao:2012pt,Cai:2014hka}, etc., are constructed from the bottom-up approach, while $\mathbf{B-L}$ can be constructed from both approaches.
Notice that new fermions are needed for anomalies cancellation of new Abelian gauge symmetry.
Of various U(1) models, ${\mathbf{B-L}}$ only requires minimal extensions of the SM with  three right-handed neutrinos, so we study its property of PT and derivative GW spectrum for simplicity. 
There are usually two types of ${\mathbf{B-L}}$ relating to the pattern of symmetry breaking:  one electroweak singlet triggered and  two electroweak singlets scalar triggered $\mathbf{B-L}$ breaking. 
We list in table.~\ref{table1}  patterns of $\mathbf{B-L}$, particle contents  as well as  their charges under $\mathbf{B-L}$, where $N_R$ represents  right-handed neutrino, $\Phi$ and $\Delta$ are electroweak singlet scalars, respectively.   
In this paper we assume $\Phi$, $\Delta$ and $Z^\prime$ are much heavier than the electroweak scale, such that the PT relating to new Abelian symmetry and electroweak symmetries breaking  can be treated separately.

\begin{table}[t]
\centering
\begin{tabular}{ccccccccccc}
\hline \hline scenario & Abelian symmetries  & $Q_L$ &$\ell_L$ & $U_R$ & $D_R$ &$E_R$& $N_R$ & $H $ & $\Phi$ & $\Delta$ \\
\hline
(a)& ${\mathbf{B-L}}$ &  ${1/3}$ &-1 &${1/ 3}$ &${1/ 3}$&-1 &-1 & 0 & 2 & $$  \\ 
\hline
(b) &${\mathbf{B-L}}$  &  ${1/3}$ &-1 &${1/ 3}$ &${1/ 3}$&-1 &-1 & 0 & 2 & 1  \\ 
\hline
\hline
\end{tabular}
\caption{ Quantum numbers of  fields under the $U(1)_{\rm B-L}^{}$,  where $\Phi$ and $\Delta$ is an electroweak  scalar singlet.  }\label{table1}
\end{table}

\subsection{Model (a)}
The Higgs potential for the  scenario  (a) of $U(1)_{\mathbf{B-L}}$ can be written as
\begin{eqnarray}
V_{0}^{(a)}= -\mu^2_{\Phi} \Phi^\dagger \Phi  + \kappa (\Phi^\dagger \Phi)^2 ,
\end{eqnarray}
where $\Phi= (\phi+ i G_\Phi + v_\Phi)/ \sqrt{2}$, with $v_\Phi$ the vacuum expectation value (VEV) of 
$\Phi$. The two parameters $\mu_{\phi}^2 $ and $\kappa$ can be replaced by the physical parameters $v_{\phi}$ and  $m_{\phi}$, $  \mu_{\phi}^2 =  {m_{\phi}^2}/{2},~  \kappa = {m_{\phi}^2}/{2 v_{\phi}^2} $.
In addition, Yukawa interactions  of  $N_R^{} $ are  
\begin{equation}
{\cal L}_{\mathbf{Y}} \sim y_N^{} \overline{N_R^C} \Phi N_R^{} + y_N^{} \overline{\ell_L^{} } \tilde{H} N_R^{} + {\rm h.c.} ,
\end{equation}
where $y_N^{}$ is $3\times 3$ symmetric Yukawa coupling matrix. 
The first term generates Majorana masses for right-handed neutrinos as $\Phi$ gets non-zero VEV.
The tiny but non-zero active neutrino masses arise from the type-I seesaw mechanism~\cite{Minkowski:1977sc}.

To study properties of the PT, one needs the effective potential at the finite temperature in terms of background field $\phi$, 
\begin{eqnarray}
V_{\rm eff}^{} &=& V_0^{} + V^{}_{\rm CW} + V_{T }^{} + V_{\rm Daisy}^{}   \nonumber \\
&=& - {1\over 2 } \mu_\Phi^2 \phi^2 + {1\over 4} \kappa \phi^4 + {1 \over 64 \pi^2 } \sum_i (-1)^{2s_i } n_i m_i^4(\phi)  \left( \log {m_i^2 (\phi ) \over \mu^2 }  -C_i \right) \nonumber \\
&+& {T^4 \over 2\pi^2 } \left\{   \sum_{i \in B } n_i J_B\left [{m_i^2 (\phi )\over T^2 } \right]- \sum_{j \in F } n_j J_F\left [{m_j^2 (\phi )\over T^2 } \right] \right \} \nonumber \\
&+& {T \over 12 \pi } \sum_i n_i \left\{  \left[ m_i^2 (\phi)\right] ^{3/2} - \left[m_i ^2( \phi) + \Pi_i (T) \right]^{3/2} \right \}  , \label{effective}
\label{effectivepotential}
\end{eqnarray}
where $V_0^{}$ is $V_0^{(a)} $ in terms of background field, $V^{}_{\rm CW}$ known as the Coleman-Weinberg potential at the zero temperature,  contains one-loop contributions to the effective potential at the zero temperature, $V_{T}$ and $V_{\rm Daisy}$ include the one-loop  and the bosonic ring contributions at the finite temperature, $n_i$ and $s_i$ are the number of degrees of freedom and the spin of the $i$-th particle, $C_i$ equals to $5/6$ for gauge bosons and $3/2$ for scalars and fermions.
Eq. (\ref{effectivepotential})  is derived in the Landau gauge. 
It should be noted that the effective potential is gauge dependent and a gauge invariant treatment of the effective potential is still unknown. 
We refer the reader to Ref.~\cite{Patel:2011th} for a gauge independent approach to the electroweak PT.
Thermal masses of scalar singlet $\phi$ and gauge boson $Z^\prime$ are given by 
\begin{eqnarray}
&&\Pi_\phi^{(a)} = \left( { g_{\mathbf{B-L}}^2 \over 2 }+ {\kappa \over 3}+ {y_N^2 \over 8} \right) T^2 \; , \\
&&\Pi_{Z^\prime}^{(a)} = {5 \over 3 } g_{\mathbf{B-L}}^2 T^2  \; ,
\end{eqnarray}
where $g_{\mathbf{B-L}}$ is the gauge coupling of $U(1)_{\mathbf{B-L}}$.
We list in the Table.~\ref{mass}  the field dependent masses of various particles.
\begin{table}[t]
\centering
\begin{tabular}{cc|cc}
\hline \hline 
scenario (a) &  & sceinario (b) \\
\hline
fields & masses & fields & masses \\
\hline
$\phi$ & $-\mu_\Phi^2 + 3 \kappa \phi^2$  &  $\phi$ & $-\mu_\Phi^2 + 3 \kappa \phi^2 + {1\over 2 }\kappa_2 \delta^2 $    \\
$\chi $ & $-\mu_\Phi^2 + ~\kappa \phi^2 $   & $\chi $ & $-\mu_\Phi^2 + ~\kappa \phi^2+ {1\over 2 }\kappa_2 \delta^2  $ \\
$N$ &  $y_{\mathbf{N}}^2 \phi^2 $  &  $N$ &  $y_{\mathbf{N}}^2 \phi^2 $   \\
$Z^\prime$  & $4 g_{\mathbf{B-L}}^2 \phi^2$ &$Z^\prime$  & $ g_{\mathbf{B-L}}^2(4 \phi^2+\delta^2)$ \\
&& $\delta$  & $-\mu_\Delta^2 +3\kappa_1 \delta^2+ {1\over 2 }\kappa_2 \phi^2  $  \\
&& $\chi^\prime$  & $-\mu_\Delta^2 +\kappa_1 \delta^2 + {1\over 2 }\kappa_2 \phi^2 $  \\
\hline
\hline
\end{tabular}
\caption{ Field-dependent masses of various particles..  }\label{mass}
\end{table}
One can see from Eq. (\ref{effectivepotential}) that the cubic term in the effective potential mainly come from the loop contribution of $Z^\prime$, such that there is strong correlation between the collider constraints on the $g_{\mathbf{B-L}}$, $m_{Z'}$ and the strength of the PT. 

\subsection{Model (b)}
The  correlation of $Z^\prime$ with the PT can be loosed in the scenario (b), where an extra scalar singlet, $\Delta\equiv ( \delta+ v_\Delta + i \chi^\prime)/\sqrt{2}$, is included.
For this scenario, the tree-level potential can be written as 
\begin{eqnarray}
  V_0^{(b)} = -\mu_\Phi^2 \Phi^\dagger \Phi + \kappa (\Phi^\dagger \Phi)^2 -\mu_\Delta^2 \Delta^\dagger \Delta + \kappa_1^{}  (\Delta^\dagger \Delta )^2 + \kappa_2 (\Phi^\dagger \Phi)(\Delta^\dagger \Delta) + \{ \Lambda \Delta^2  \Phi^{\dagger} + {\rm h.c. } \} ,\label{potentialb}
\label{eq:v-modelb}
\end{eqnarray}
where $\Lambda$ is a coupling with energy scale. 
$\mu_{\Phi}^2$ and $\mu_{\Delta}^2$ can be replaced  with $v_{\phi}$ and $v_{\delta}$ via the tadpole conditions 
\begin{align}
& \mu_{\phi}^2 = \frac{1}{2} \kappa_2 v_{\delta}^2 + \kappa v_{\phi}^2 + \frac{\Lambda v_{\delta}^2}{\sqrt{2} v_{\phi}} , \\
& \mu_{\Delta}^2 = \kappa_1 v_{\delta}^2 + \frac{1}{2} \kappa_2 v_{\phi}^2 + \sqrt{2} \Lambda v_{\phi} .
\end{align}
The mass matrix for the CP-even scalars follows,
\begin{align}
  \mathcal{M}_{\phi,\delta}^2 = 
\left(
\begin{array}{cc}
2 v_{\phi}^2 \kappa - \frac{v_{\delta}^2 \Lambda}{\sqrt{2} v_{\phi}} &  v_{\delta}(v_{\phi} \kappa_2 + \sqrt{2} \Lambda) \\
v_{\delta}(v_{\phi} \kappa_2 + \sqrt{2} \Lambda)   & 2 v_{\delta}^2 \kappa_1 ,
\end{array}
\right ),
\end{align}
which can be diagonalized by a $2\times 2$ orthogonal matrix parametrized by a rotation angle $\theta$, 
\begin{equation}
s_1 = c_{\theta} \phi + s_{\theta} \delta, \quad \quad \quad s_2 = -s_{\theta} \phi + c_{\theta} \delta,
\end{equation}
where $s_{1,2}$ are mass eigenstates with mass eigenvalues $m_{s_1}$ and $m_{s_2}$ respectively. 
Three quartic couplings can now be  written in term of  physical parameters,
\begin{align}
& \kappa_1 = \frac{m_{s_1}^2 s_{\theta}^2 + m_{s_2}^2 c_{\theta}^2}{2 v_{\theta}^2}, \\
& \kappa_2 = \frac{s_{\theta} c_{\theta} (m_{s_1}^2 - m_{s_2}^2) - \sqrt{2} \Lambda v_{\delta}}{v_{\delta} v_{\phi}}, \\
& \kappa \ = \frac{2 m_{s_1}^2 c_{\theta}^2 v_{\phi} + 2 m_{s_2}^2 s_{\theta}^2 v_{\phi} + \sqrt{2} \Lambda v_{\delta}^2}{4 v_{\phi}^3}.
\end{align}
For the CP-odd scalars, their mass matrix is given by
\begin{eqnarray}
  \mathcal{M}^2_{G_{\phi},\chi^{\prime}} = 
- \frac{\Lambda}{\sqrt{2} v_{\phi}}
\left(
\begin{array}{cc}
v_{\delta}^2            &   -2 v_{\delta} v_{\phi} \\
-2 v_{\delta} v_{\phi}  &   4 v_{\phi}^2
\end{array}
\right)  .
\end{eqnarray}
It can be diagonalized by a rotation matrix with angle  $\theta^{\prime} = \arctan [v_{\delta}/(2 v_{\phi})]$ and gives the following mass eigenstates 
\begin{equation}
G_{Z^{\prime}} 
= c_{\theta^{\prime}} G_{\phi} + s_{\theta^{\prime}} \chi^{\prime}, 
\quad \quad \quad 
A = -s_{\theta^{\prime}} G_{\phi} + c_{\theta^{\prime}} \chi^{\prime},
\end{equation}
where $G_{Z^{\prime}}$ is the Goldstone boson and $A$ is the physical CP-odd scalar with its mass given by $m_A^2 = -\Lambda (v_\delta^2 + 4 v_\phi^2 ) /\sqrt{2} v_\phi$, which implies $\Lambda < 0$. 
The physical parameters in this scenario are then
\begin{equation}
v_{\phi}, \quad v_{\delta}, \quad m_{s_1}, \quad m_{s_2}, \quad \theta, \quad \Lambda.
\end{equation}
The effective potential of the scenario (b) has the same form as Eq. (\ref{effective}) up to the following replacements: $(a)\to (b)$, $m_{i} (\phi) \to m_i ( \phi , \delta)$. The field dependent masses are tabulated in the second column of Table.~\ref{mass}, while the thermal masses of the various fields are given below,
\begin{eqnarray}
&&\Pi_\phi^{(b)} = \left( { g_{\mathbf{B-L}}^2 \over 2 }+ {\kappa \over 3}+ {\kappa_2 \over 12 }+ {y_N^2 \over 8} \right) T^2 \; , \\
&&\Pi_\delta^{(b)} = \left( { g_{\mathbf{B-L}}^2 \over 4 }+ {\kappa_1 \over 3}+ {\kappa_1 \over 12 } \right) T^2 \; , \\
&&\Pi_{Z^\prime}^{(b)} = {7 \over 4 } g_{\mathbf{B-L}}^2 T^2  \; .
\end{eqnarray}
With these inputs, the phase history can be analyzed. 
A particular advantage of model(b) is that there is a cubic term in Eq. (\ref{potentialb}) at the tree-level, which  can generate a barrier between the broken and symmetric phases without the aid of loop corrections. 
As a result it is easier to get a first oder PT for this scenario, compared with model(a) where the barrier is provided by $Z'$ from loop corrections.

We now address collider constraints on the $Z^\prime $ mass.  
A heavy $Z^\prime$ with SM $Z$ couplings to fermions was searched at the LHC in the dilepton channel,  which is excluded at the 95\% CL for $M_{Z^\prime} <2.9$ TeV from the ATLAS~\cite{Aad:2014cka} and for $M_{Z^\prime}< 2.79$ TeV from the CMS~\cite{Khachatryan:2014fba}. 
The measurement of $e^+ e^- \to f \bar f$ above the Z-pole at the LEP-II puts  lower bound on $M_{\rm Z^\prime } /g_{\rm new}$, which is about $6$ TeV~\cite{Carena:2004xs}. 
Further constraint is given by the ATLAS collaboration~\cite{Aaboud:2017buh}  with $36.1~fb^{-1}$ of proton-proton collision data collected at $\sqrt{s}$= 13 TeV, which has $M_{Z_\mathbf{B-L}} >4.2~\text{TeV}$. 
We keep these constraints in studying PTs of these models.

\section{Gravitational wave signals}\label{sec:GWs}
For parameter settings of these two models that can give a first order phase transition, there will be 
gravitational waves generated, mainly coming from three processes: 
bubble collisions, sound waves in the plasma and Magnetohydrodynamic 
turbulence(see Ref.~\cite{Caprini:2015zlo,Cai:2017cbj,Weir:2017wfa} for recent reviews).  
The total energy spectrum can be written approximately as the sum of these three contributions:
\begin{equation}
  \Omega_{\text{GW}}h^{2} \simeq \Omega_{col}h^{2}+\Omega_{sw}h^{2}+\Omega_{turb}h^{2} ,
\end{equation}
where the Hubble constant is defined following the conventional way 
$H = 100h \  \textrm{km} \textrm{s}^{-1} \textrm{Mpc}^{-1}  $. 
The energy spectrums depend on three important input parameters for each 
specific particle physics model: the bubble wall velocity($\equiv v_w$), 
\begin{equation}
\alpha=\left. \frac{\Delta\rho}{\pi^{2}g_{\ast}T^{4}/30}\right|_{T = T_{n}}, \quad 
\text{and} \quad  \beta= H_n T_{n} \left. \frac{d(S_{3}/T)}{dT}\right|_{T =T_{n}} ,
\end{equation}
where $ \Delta \rho $ is the difference of energy density between the false 
and true vacua, $g_{\ast}$ is the number of relativistic degrees of freedom and $ H_{n} $ is the Hubble constant 
evaluated at the nucleation temperature $T_{n}$, which
corresponds approximately to the temperature when $S_3(T)/T = 140$~\cite{Apreda:2001us}.
The parameter $\alpha$ characterizes the strength of the PT while $\beta$ denotes roughly
the inverse time duration of the PT. 
With these parameters solved numerically, one can obtain the energy spectrum of the gravitational waves for three sources.

Firstly for the GW from the bubble collision, it can be calculated using the 
envelop approximation~\cite{Kosowsky:1991ua,Kosowsky:1992rz,Kosowsky:1992vn} either by numerical simulations~\cite{Huber:2008hg} or by a recent analytical 
approximation~\cite{Jinno:2016vai}. 
Both results can be summarized in the following form,
\begin{equation}
\Omega_{col}h^{2}=1.67\times10^{-5} \Delta (v_w) \left( \frac{H_{n}}{\beta}\right)^{2}
\left(\frac{\kappa_{\phi} \alpha}{1+\alpha} \right)^{2} \left( \frac{100}{g_{\ast}}\right)^{1/3} 
S_{\text{env}}(f) .
\end{equation}
Here $\kappa_{\phi}$ is the fraction of latent heat transferred to the scalar field gradient, $\Delta (v_w)$ is a numerical factor
and $S_{\text{env}}$ captures the spectral shape dependence. 
The two different treatments by  Ref.~\cite{Jinno:2016vai} and Ref.~\cite{Huber:2008hg} lead to slightly different results on the $\Delta (v_w)$ and $S_{\rm env}$. 
We adopt here the results from the numerical simulation,
\begin{equation}
\Delta (v_w) = \frac{0.48 v_w^3}{1+5.3 v_w^2 + 5 v_w^4} , \quad \quad
S_{\text{env}} =  { 3.8 (f /f_{\rm env})^{2.8} \over 1+ 2.8 (f/f_{\rm env})^{3.8}}  ,
\end{equation}
with $ f_{\textrm{env}} $ the peak frequency at present time given by,
\begin{equation}
f_{\textrm{env}}=16.5\times10^{-6}\left(\frac{f_{\ast}}{\beta} \right)\left(\frac{\beta}{H_{n}} \right)  \left( \frac{T_{n}}{100\textrm{GeV}} \right) \left( \frac{g_{\ast}}{100}\right)^{1/6} \textrm{Hz} ,
\end{equation}
which is the redshifted frequency of the peak frequency, $f_{\ast}$, at the time of the PT,
\begin{equation}
f_{\ast} = \frac{0.62}{1.8-0.1 v_w + v_w^2} .
\end{equation}
For the spectral shape  $S_{\text{env}}$, the analytical treatment in Ref.~\cite{Jinno:2016vai} shows the correct behavior 
for low frequency $S_{\text{env}} \propto f^3$ required by causality~\cite{Caprini:2009fx} while 
the result from the numerical simulations differs from this one in a minor way. 
According to a more recent paper~\cite{Bodeker:2017cim}, in which the runaway conclusion~\cite{Bodeker:2009qy} of the bubble expansion is ruled-out, the energy deposited
in the scalar field is negligible and should be neglected in GW calculations. 
We therefore neglect the contribution of bubble collision due to the smallness of $\kappa_\phi$. 
 
Secondly, the bulk motion of the fluid in the form of sound wave are produced after the bubble collisions.
It also generates GWs and the energy spectrum has been simulated, with~\cite{Hindmarsh:2015qta},
\begin{equation}
\Omega_{\textrm{sw}}h^{2}=2.65\times10^{-6}\left( \frac{H_{n}}{\beta}\right)^{2}\left(\frac{\kappa_{v} \alpha}{1+\alpha} \right)^{2} \left( \frac{100}{g_{\ast}}\right)^{1/3}v_{w} \left(\frac{f}{f_{sw}} \right)^{3} \left( \frac{7}{4+3(f/f_{\textrm{sw}})^{2}} \right) ^{7/2} \ .
\end{equation}
Here $f_{\text{sw}}$ is the peak frequency at current time redshifted from the one at the phase transition:
$2 \beta/(\sqrt{3}v_w)$, then
 \begin{equation}
f_{\textrm{sw}}=1.9\times10^{-5}\frac{1}{v_{w}}\left(\frac{\beta}{H_{n}} \right) \left( \frac{T_{n}}{100\textrm{GeV}} \right) \left( \frac{g_{\ast}}{100}\right)^{1/6} \textrm{Hz} .
\end{equation}
Similar to $\kappa_{\phi}$, the factor $ \kappa_{v} $ is the fraction of latent heat transformed into the 
bulk motion of the fluid. 
We use the method summarized in Ref.~\cite{Espinosa:2010hh} to calculate $\kappa_v$
as a function of ($\alpha$, $v_w$) and note that a fitted approximate formula is given in Ref.~\cite{Espinosa:2010hh}. 
We also note that  a more recent numerical simulation by the same collaboration~\cite{Hindmarsh:2017gnf} gives a slightly
enhanced $\Omega_{\text{sw}} h^2$ and a slightly reduced peak frequency $f_{\text{sw}}$.

Finally the plasma at the time of phase transition is fully ionized and the resulting MHD turbulence can 
give another source of GWs. 
Neglecting a possible helical component~\cite{Kahniashvili:2008pe}, the 
generated GW spectrum can be modeled in a similar way~\cite{Caprini:2009yp,Binetruy:2012ze},
\begin{equation}
\Omega_{\textrm{turb}}h^{2}=3.35\times10^{-4}\left( \frac{H_{n}}{\beta}\right)^{2}\left(\frac{\kappa_{turb} \alpha}{1+\alpha} \right)^{3/2} \left( \frac{100}{g_{\ast}}\right)^{1/3} v_{w}  \frac{(f/f_{\textrm{turb}})^{3}}{[1+(f/f_{\textrm{turb}})]^{11/3}(1+8\pi f/h_{\ast})} ,
\end{equation}
with the peak frequency $f_{turb}$ given by,
\begin{equation}
f_{\textrm{turb}}=2.7\times10^{-5}\frac{1}{v_{w}}\left(\frac{\beta}{H_{n}} \right) \left( \frac{T_{n}}{100\textrm{GeV}} \right) \left( \frac{g_{\ast}}{100}\right)^{1/6} \textrm{Hz} .
\end{equation}
We need to know  the factor $\kappa_{\text{turb}}$ which is the fraction of latent heat transferred to MHD 
turbulence. 
The precise value is still undetermined and a recent numerical simulation shows that  $\kappa_{\rm turb}$ can be  parametrized as $\kappa_{\text{turb}}\approx \epsilon \kappa_{v} $, where the numerical 
factor $\epsilon$ varying roughly between $5 \sim 10\%$~\cite{Hindmarsh:2015qta}. 
Here we take tentatively $\epsilon = 0.1$.
\begin{figure}
\includegraphics[width=0.3\textwidth]{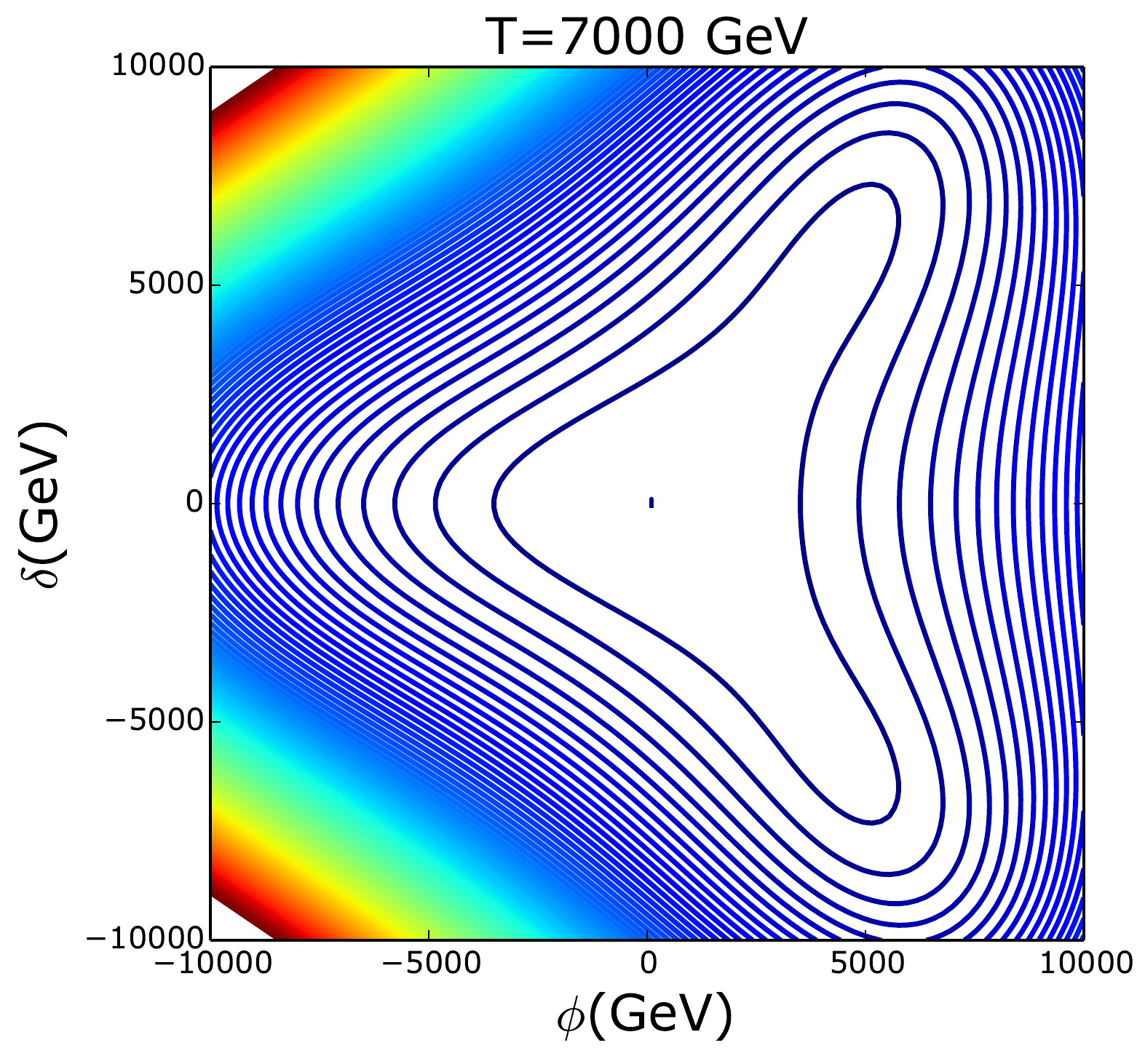}
\includegraphics[width=0.3\textwidth]{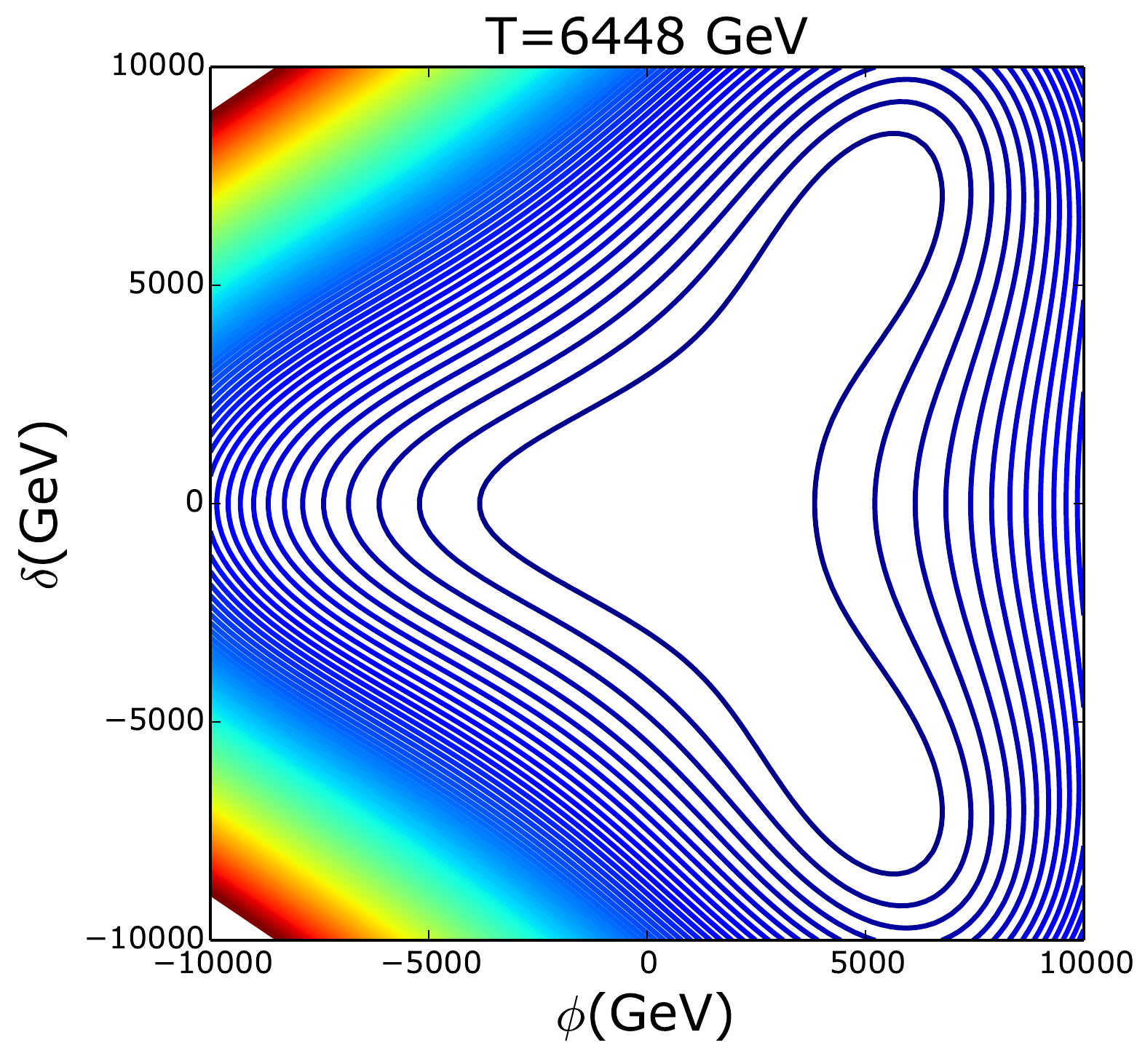}
\includegraphics[width=0.3\textwidth]{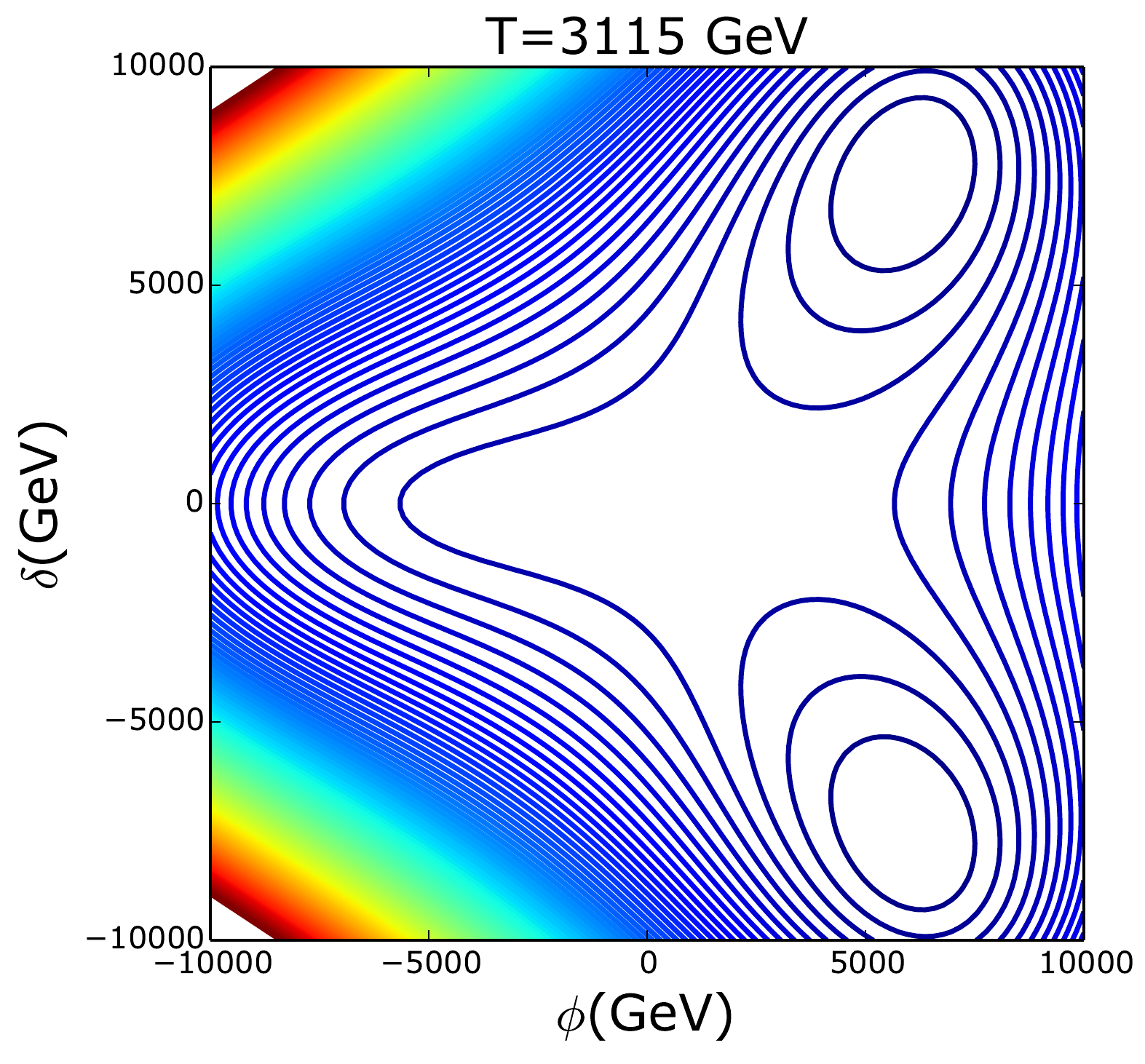}
\caption{
\label{fig:modelb-v}
The contours of the effective potential of model(b) at three typical temperatures, with blue
lines for lower values and red for higher values. The left figure is at a temperature higher than 
$T_C \approx 6448 \text{GeV}$, the middle 
one is at $T_C$ and the right figure is at $T_n \approx 3115\text{GeV}$.
The benchmark parameters are chosen as: $v_{\phi}=4637 \text{GeV}$, $v_{\delta} = 1902 \text{GeV}$,
$\theta = 0.128$, $m_{s_1} = 2400 \text{GeV}$, $m_{s_2} = 1236 \text{GeV}$ and $\Lambda = -2143\text{GeV}$.
}
\end{figure}

For detection of the GWs, one needs to compare these spectrums with the sensitivity curve of each 
detector. 
The LISA detector~\cite{Audley:2017drz} is currently the most mature experiment and the recently 
finished LISA pathfinder has confirmed its design goals. 
We therefore consider the sensitivities of the four LISA configurations 
N2A5M5L6(C1), N2A1M5L6(C2), N2A2M5L4(C3), N1A1M2L4(C4) presented in 
Ref.~\cite{Klein:2015hvg,Caprini:2015zlo}, which include 
the instrumental noise of the LISA detector obtained using the detector simulation package 
LISACode~\cite{Petiteau:2008zz} as well as the astrophysical foreground from the compact white dwarf binaries 
in our Galaxy. 
We also consider the discovery prospect of several other proposed experiments:
the Advanced Laser Interferometer Antenna (ALIA)~\cite{Gong:2014mca}\footnote{It is now renamed as Taiji.}, the Big Bang Observer(BBO), 
the DECi-hertz Interferometer Gravitational wave 
Observatory(DECIGO)~\footnote{The ALIA, BBO and DECIGO sensitivity data are taken from the 
website~\url{http://rhcole.com/apps/GWplotter/}} and Ultimate-DECIGO~\cite{Kudoh:2005as}. 
%
\begin{figure}
\centering
\includegraphics[width=0.4\textwidth]{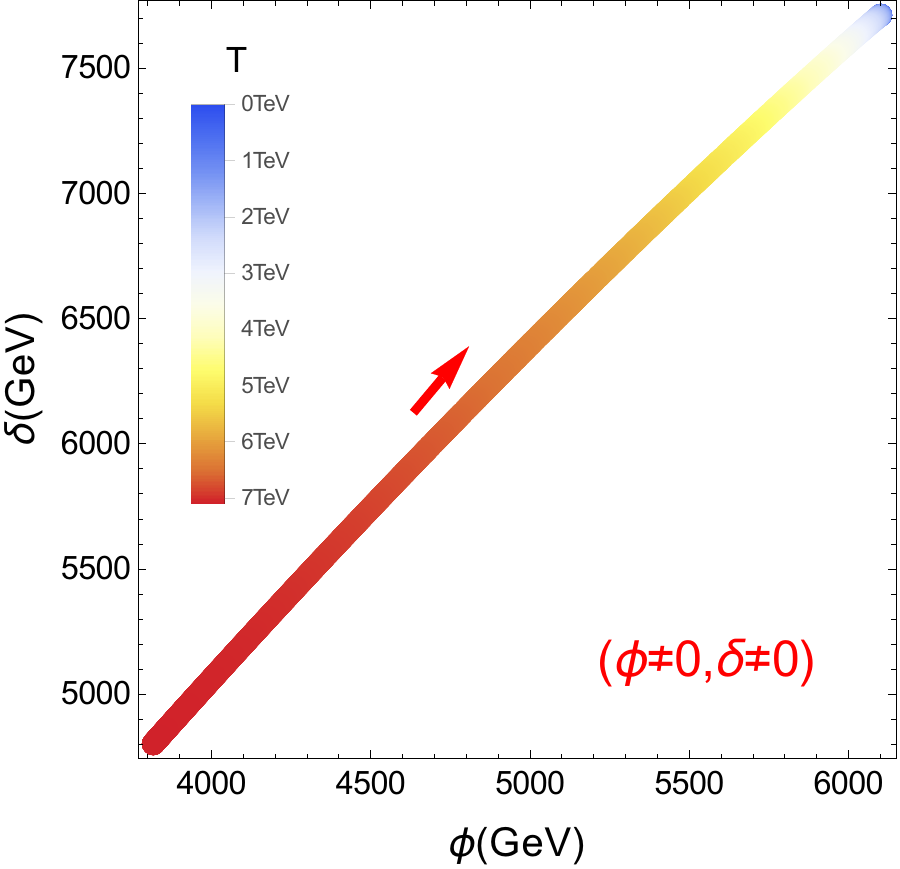}
\caption{ 
\label{fig:phase}
The tracks of the minimum $(\phi \neq 0, \delta \neq 0)$ in the $(\phi, \delta)$
plane with the colors showing the value of temperature, which can be read from the colormap on the left.
}
\end{figure}

We implement two $\mathbf{B-L}$ models in CosmoTransitions~\cite{Wainwright:2011kj} which traces the phase 
history of each model, locates the critical temperature $T_C$ and gives the bounce solutions to obtain
the bubble nucleation temperature $T_n$. 
We then use these outputs to calculate the GW energy  spectrums and compare them with the listed detector sensitivities.

From an extensive scan over the parameter space of model(a) at the mass scale of 
$\mathcal{O}(\text{TeV})$, we find that a first 
order PT can occur for a significant proportion of their parameter spaces. However, the resulting GW signals 
are generally too weak to be discovered where the most optimistic case can marginally be reached by 
the Ultimate-DECIGO. 
This is due to the relatively large values of $\beta$ and small values of $\alpha$ 
obtained, aside from the enhanced $\mathcal{O}(\text{TeV})$ temperature, which reduce the magnitude of 
GW energy spectrum as well as pushing the peak frequency to higher 
values. 
On the other hand, for the parameter space at the electroweak scale, the GWs can generally be 
reached by most detectors, which is however ruled by collider searches of $Z^\prime$.
%
\begin{figure}[t]
\centering
\includegraphics[width=0.4\textwidth]{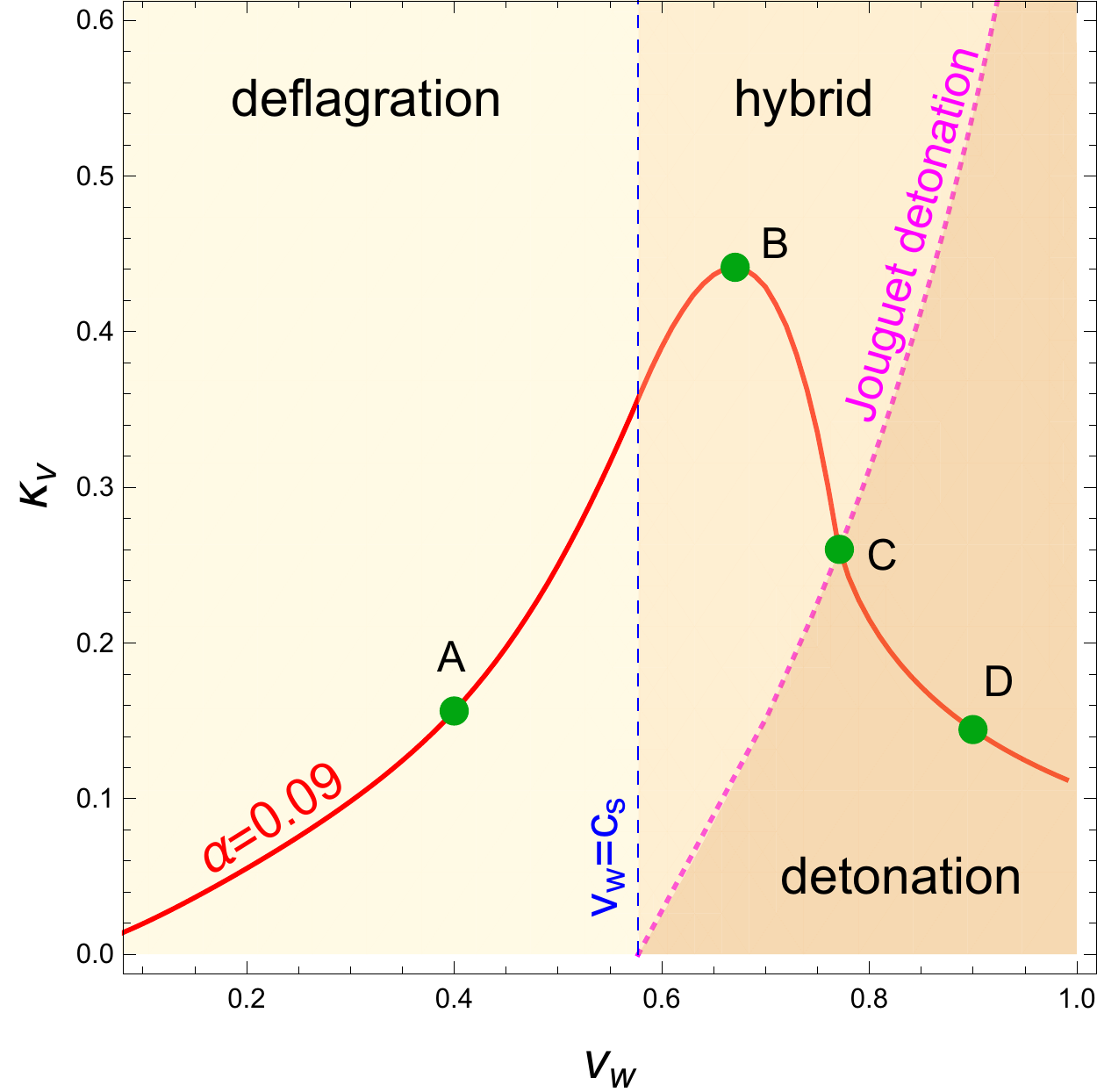}
\quad \quad
\includegraphics[width=0.4\textwidth]{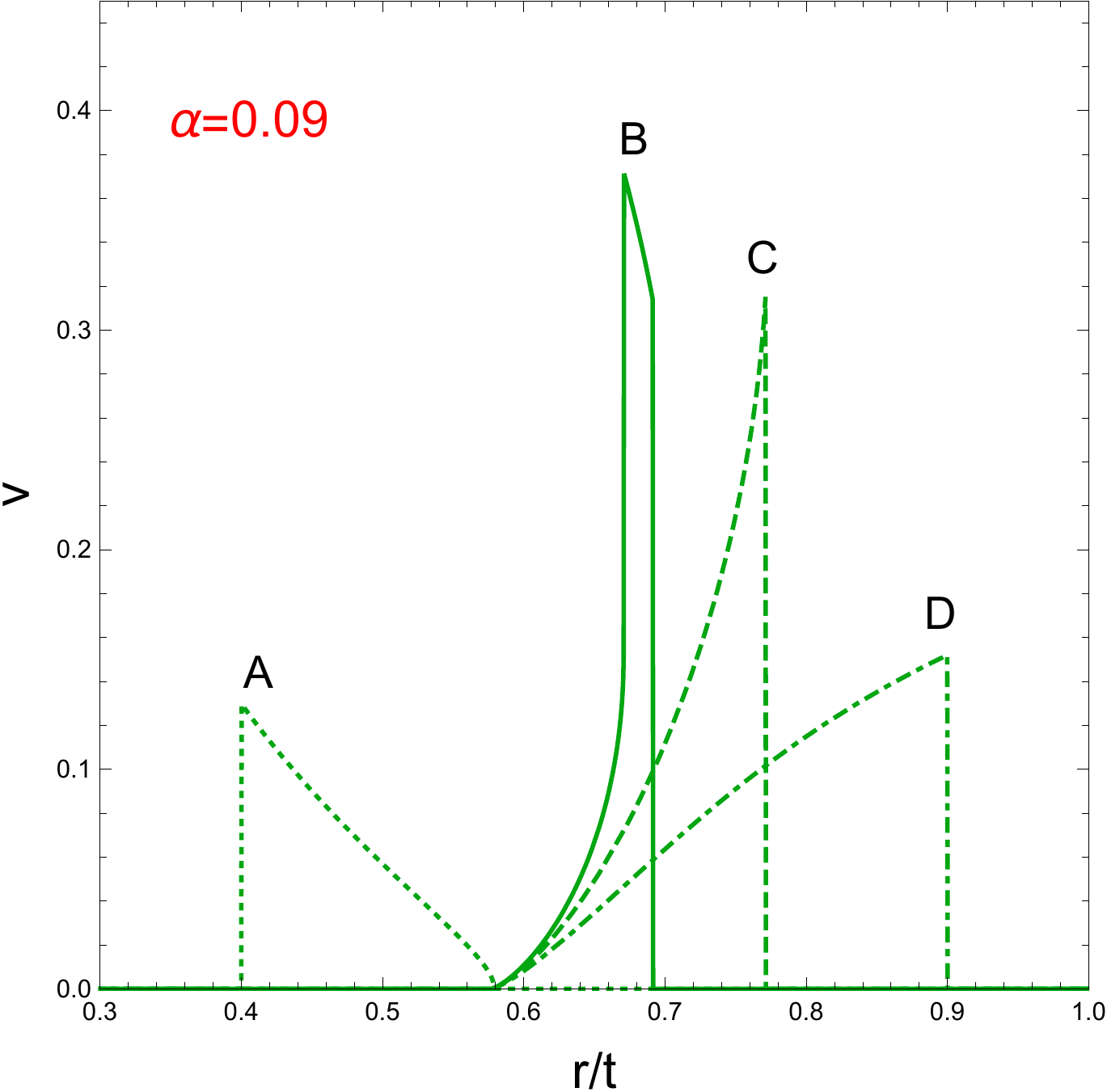}
\caption{ 
\label{fig:kappa}
Left panel: The red line shows the fraction of latent heat transferred to the bulk motion of the plasma 
$\kappa_v$ when the bubble wall velocity is varied for $\alpha=0.09$, which is derived the benchmark in Fig.~\ref{fig:modelb-v}. Also plotted here are the deflagration, hybrid(supersonic deflagration) and detonation
regions characterizing the dynamics of the phase transition, separated by the blue dashed line(when $v_w$ is
equal to the sound speed of the relativistic plasma $c_s = 1/\sqrt{3}$) and the magenta dotted 
line(Jouguet detonation). Four representative cases: A, B, C and D, marked with green points, 
are chosen to calculate the GW spectra. Right panel: the velocity profile as a function of $r/t$ for 
the four representative cases of the left panel plot.
}
\end{figure}

Model(b) has a sizable parameter space where the generated GWs from PT falls within the sensitive regions of various detectors, due to the easily 
realized PT from the tree level barrier with the aid of a negative cubic term in the effective 
potential in Eq.~\ref{eq:v-modelb}. 
We show a benchmark point from this parameter space and present the  details of the PT and the GW spectrum. 
This benchmark parameter point is 
$v_{\phi}=4637 \text{GeV}$, $v_{\delta} = 1902 \text{GeV}$, $\theta = 0.128$, 
$m_{s_1} = 2400 \text{GeV}$, $m_{s_2} = 1236 \text{GeV}$ and $\Lambda = -2143\text{GeV}$.
For this case, the minima in the field space $(\phi,\delta)$ lie in the 
direction $\phi > 0$, where the cubic term in Eq.~\ref{eq:v-modelb} is negative. Due to the reflection 
symmetry $\delta \rightarrow - \delta$, this occur in a pair. 
The shape of the effective potential is 
shown as contours in Fig.~\ref{fig:modelb-v} where hot regions have larger values of $V$ while cold 
regions have smaller values. 
The leftest figure shows the shape at a relatively high temperature where 
the universe sits at its origin and the two minima in direction $\phi > 0$ are developing. 
As $T$ drops to the critical temperature $T_C \approx 6448 \text{GeV}$, these two minima become degenerate with the 
one at the origin as shown in the middle figure.
 As $T$ further drops below the critical temperature,
the broken phase begin to nucleate on the background of symmetric phase at $T_n \approx 3115 \text{GeV}$,  which corresponds to the rightest figure.
The details on the evolution of the new phase  is shown in Fig.~\ref{fig:phase} in the plane $(\phi, \delta)$ where the arrow denotes the direction of time flow and the colors show the value of temperature.
%

\begin{figure}[!ht]
\centering
\includegraphics[width=0.43\textwidth]{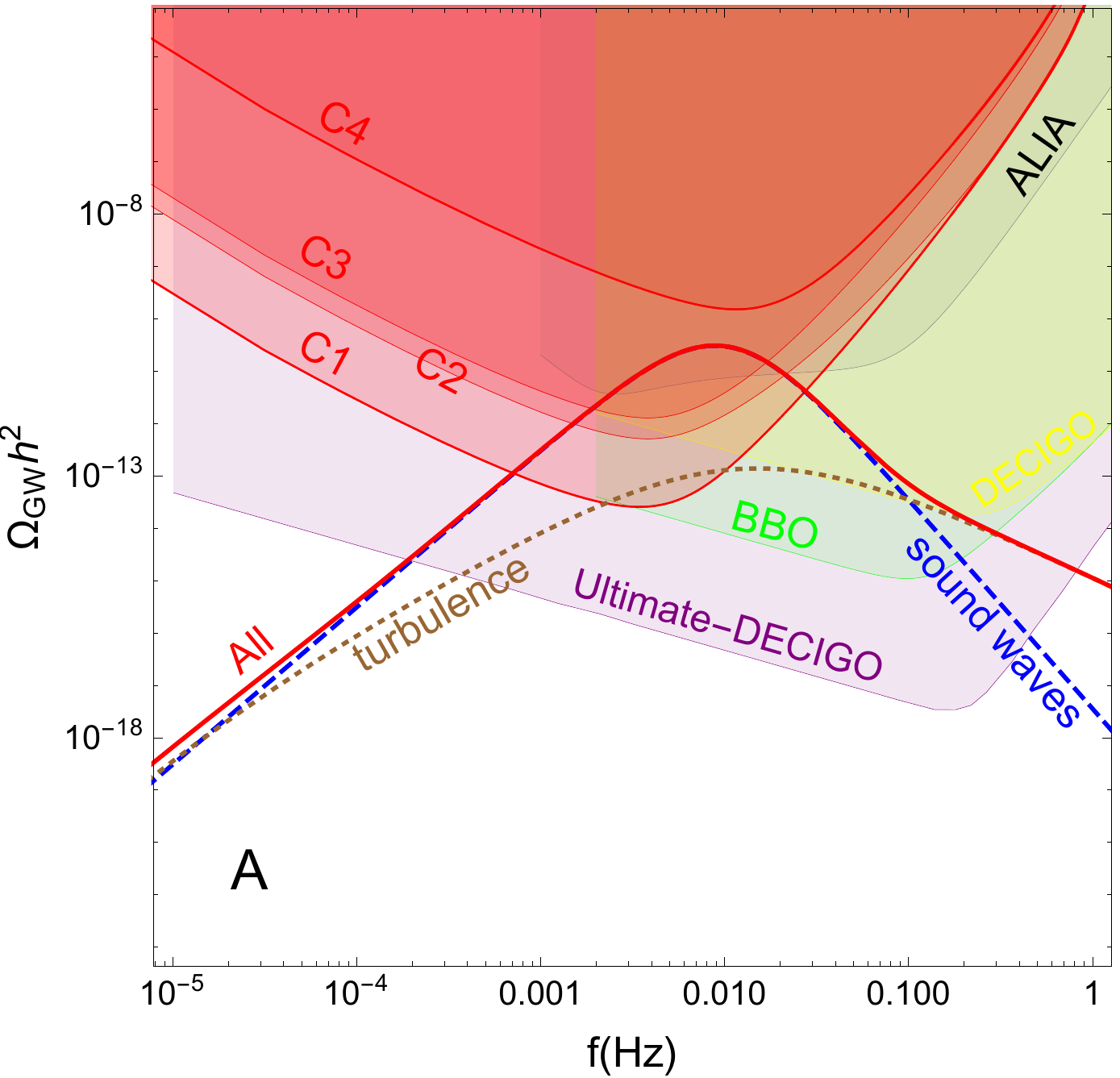}
\quad
\includegraphics[width=0.43\textwidth]{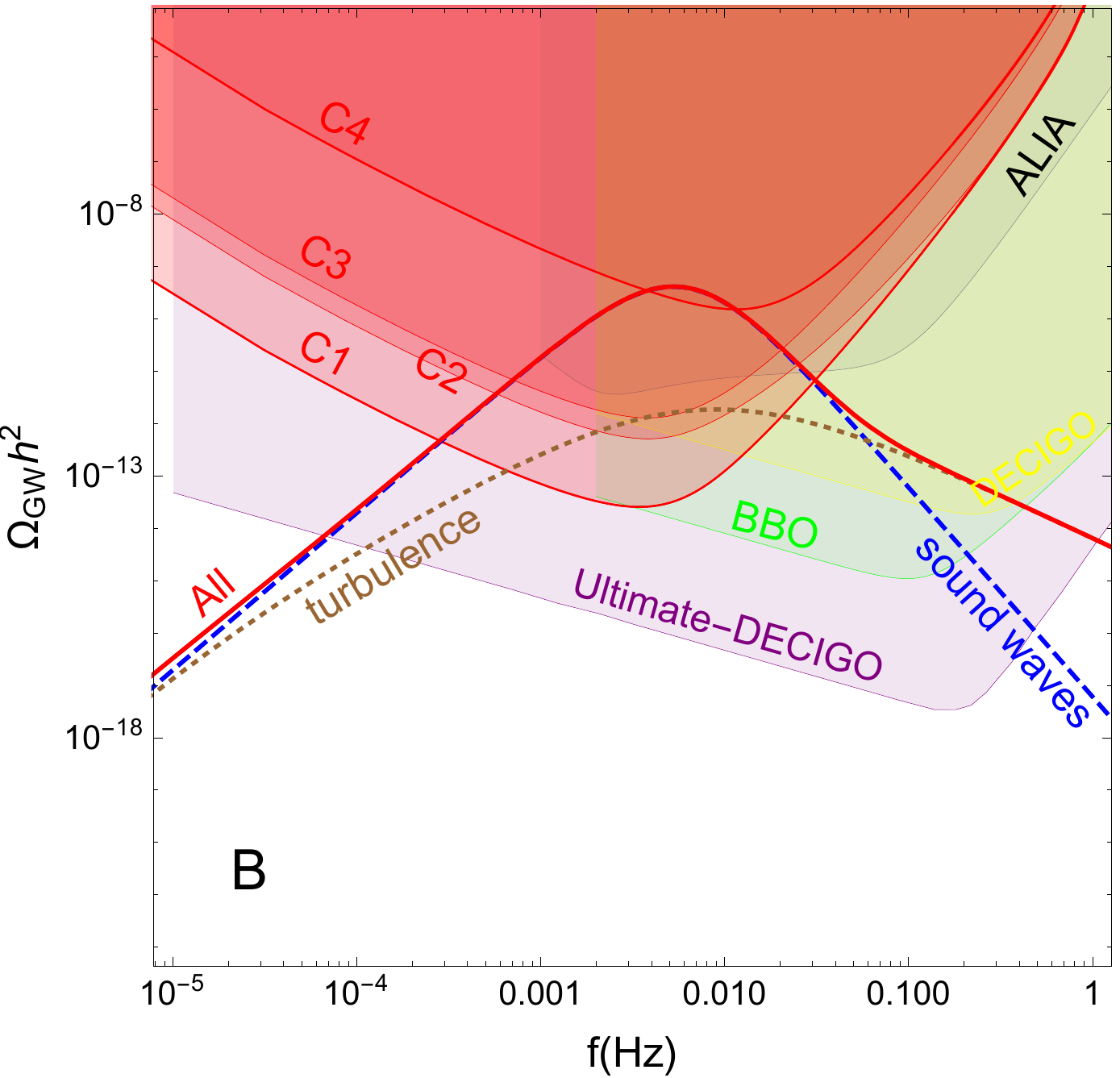}
\\
\includegraphics[width=0.43\textwidth]{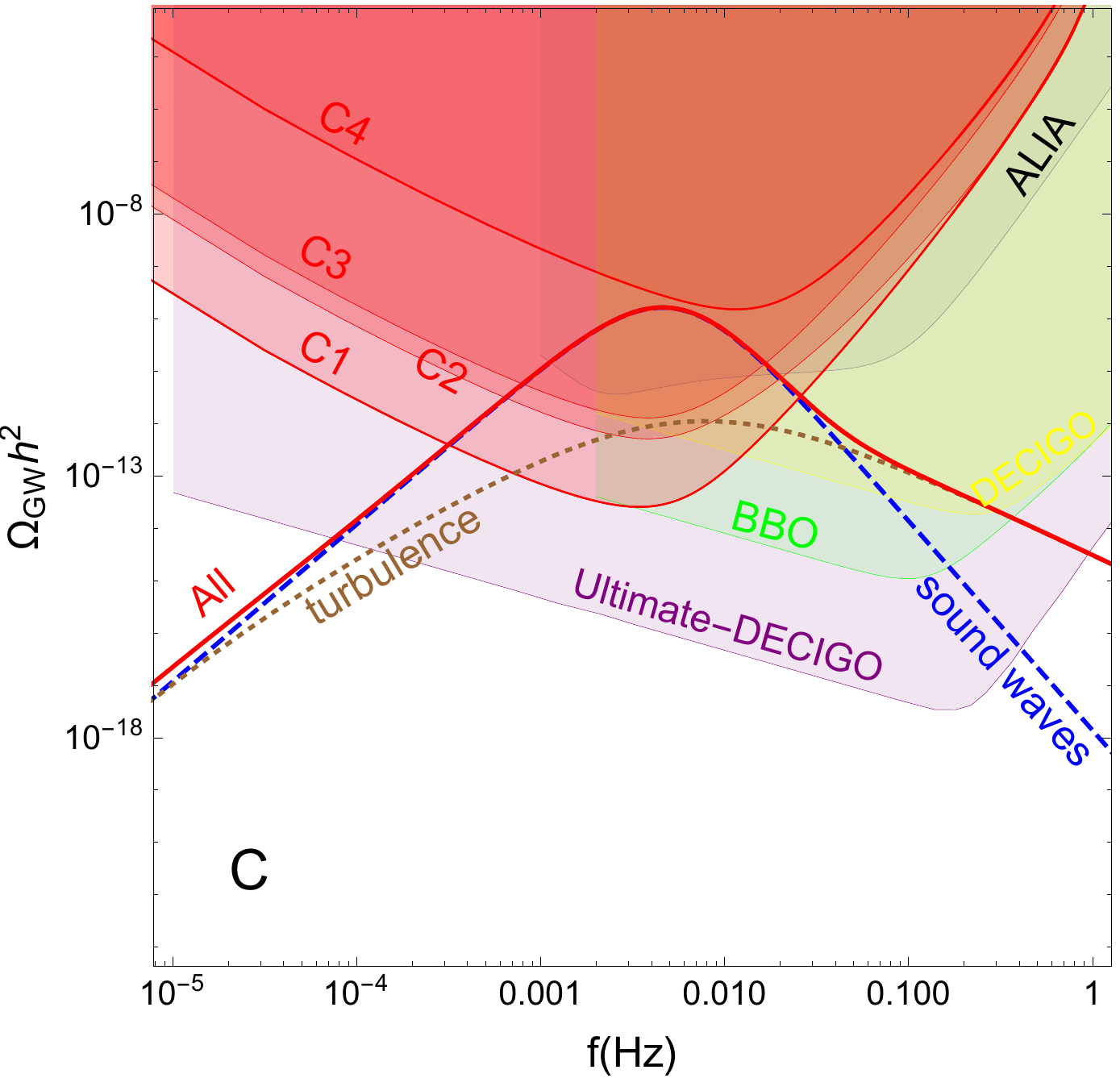}
\quad
\includegraphics[width=0.43\textwidth]{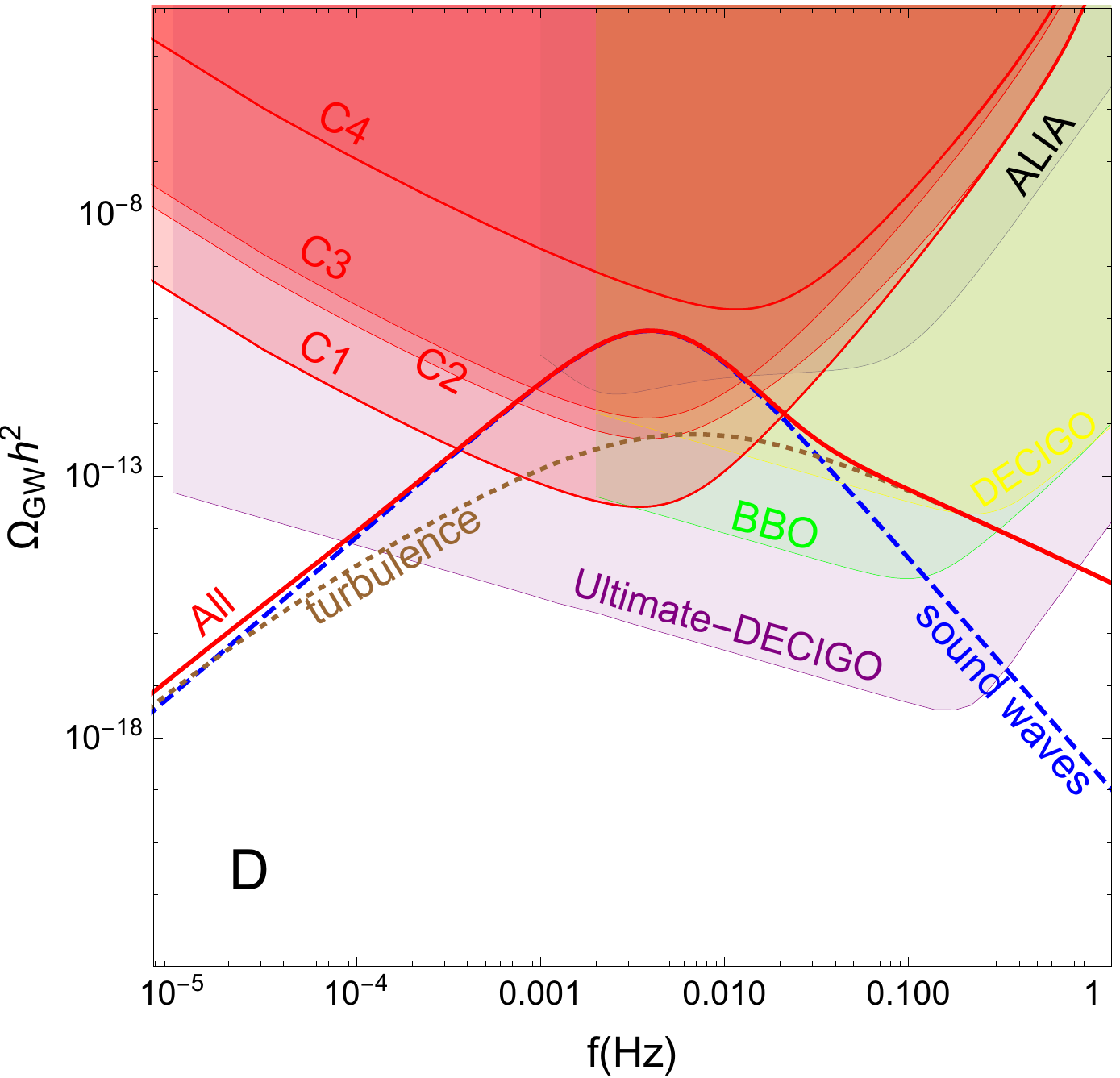}
\caption{ 
\label{fig:gw-modelb}
The GW energy spectrum  as the function of its frequency  for the benchmark in Fig.~\ref{fig:modelb-v} and four representative bubble wall velocities (the four green points A,B C and D in Fig.~\ref{fig:kappa}).
The individual contributions from sound waves and turbulence are plotted using 
blue dashed, brown dotted lines respectively, with their sum corresponding to the red solid line. 
Also plotted are the experimental sensitive regions at the top, corresponding to color-shaded regions, 
from four configurations of the LISA detector C1-C4(red), ALIA(gray), DECIGO(yellow), 
BBO(green) and Ultimate-DECIGO(purple).
}
\end{figure}

To calculate the GWs from this model, we need the input $\kappa_v$ which we calculate following 
Ref.~\cite{Espinosa:2010hh}. 
For benchmark given in Fig.~\ref{fig:modelb-v}, we find $\alpha=0.09$ and  $\kappa_v$  depends on one free parameter $v_w$. 
For different values of $v_w$, the motion of the plasma surrounding the bubble takes different forms and the value of $\kappa_v$ is shown in the left panel of 
Fig.~\ref{fig:kappa},  where representative points are selected marked as  A, B, C and D  shown as green points in the figure.  
The velocity  profiles of the plasma  is shown in the right panel of Fig.~\ref{fig:kappa} as a function of $r/t$, where $r$ is the radial distance from the 
bubble center and $t$ starts at $T_n$. 
For case A,  $v_w$ is smaller than the sound speed in the  plasma($\equiv c_s =1/\sqrt{3}$, the vertical dashed line in left panel),  and the bubble proceeds 
as deflagrations, with a velocity profile shown by the dotted lines in the right panel.
For case B, $v_w$ is larger than $c_s$, a rarefaction wave develops behind the bubble wall, yet the 
fluid has non-zero velocity ahead of the wall, corresponding to the solid lines in the right panel. 
This  falls within the hybrid region of the left panel, denoting supersonic deflagration~\cite{KurkiSuonio:1995pp}. 
For case C, $v_w$ is increased to the Jouguet detonation~\cite{Steinhardt:1981ct} (the magenta dotted line in the left panel) and 
the velocity of the fluid ahead of the wall becomes zero, corresponding to the dashed line in the right panel.
For case D, the bubble wall velocity gets larger and the expansion takes the form of detonation with the profile shown by the dot-dashed line in the right panel.

The resulting GW energy spectrums for these four  points  from 
sound waves(blue dashed) and turbulence(brown dotted) are shown in Fig.~\ref{fig:gw-modelb}, where their sum corresponds to the red solid line. 
The color-shaded regions at the top 
are the experimental sensitivity regions for the 
four LISA configurations C1-C4(red), ALIA(gray),
DECIGO(yellow), BBO(green) and Ultimate-DECIGO(purple). 
It is observed that for all four cases, the spectrum at around the peak frequency is dominated by sound waves
while turbulence becomes more important for large and small frequencies. 
The total GW spectra all fall
within the experimental sensitive regions of the LISA configurations C1, C2, C3 as well as other experiments.
For case B, corresponding to the peak of $\kappa_v$ in the left panel of Fig.~\ref{fig:kappa}, the least sensitive configuration of LISA C4 can also reach some proportion of the GW spectrum even though the 
resulting signal-to-noise ratio might be too small.

\begin{figure}
\centering
\includegraphics[width=0.6\textwidth]{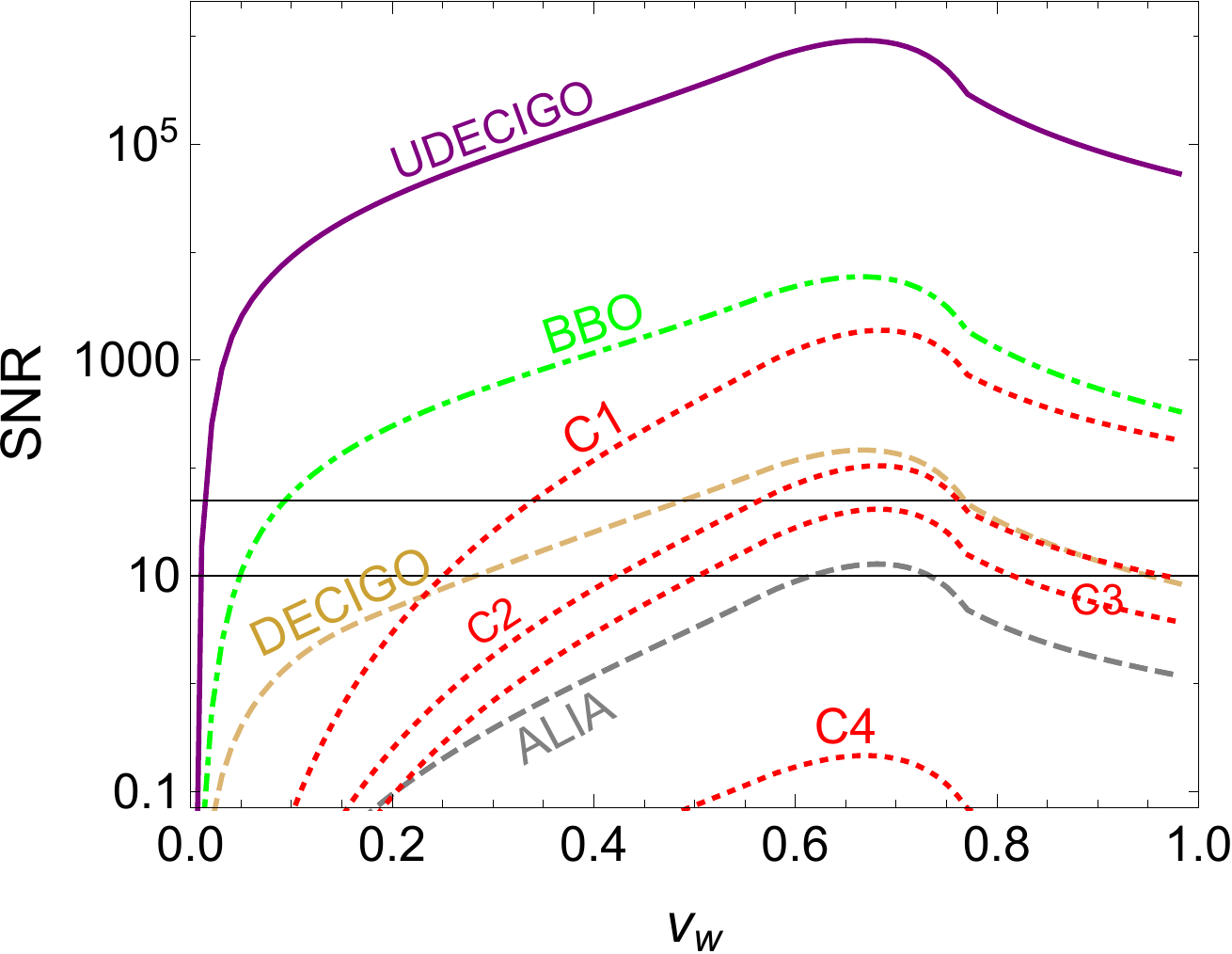}
\caption{
\label{fig:snr}
The SNR as the function of bubble wall velocity $v_w$ for the benchmark point in model(b) 
using the different experimental sensitivity inputs. 
Two black horizontal lines denote the SNR threshold
values 10 and 50 respectively.
}
\end{figure}

To assess the discovery prospect of the GWs, we quantify the detectability of the GWs using 
the signal-to-noise ratio adopted in Ref.~\cite{Caprini:2015zlo}:
\begin{eqnarray}
  \text{SNR} = \sqrt{\mathcal{T} \int_{f_{\text{min}}}^{f_{\text{max}}} df 
    \left[
      \frac{h^2 \Omega_{\text{GW}}(f)}{h^2 \Omega_{\text{exp}}(f)} 
  \right]^2} ,
\end{eqnarray}
where $h^2 \Omega_{\text{exp}}$ is the experimental sensitivity shown in Fig.~\ref{fig:gw-modelb} and 
$\mathcal{T}$ is the mission duration of the experiment in years.
With this formula, we calculate SNR as a function of $v_w$ for each experiment and show the results in Fig.~\ref{fig:snr}. 
We also show two representative SNR thresholds  $\text{SNR}_{\text{thr}} = 10, 50$ as suggested 
by Ref.~\cite{Caprini:2015zlo} with horizontal black lines for comparison. 
From this figure, we can see that 
all SNR curves have a peak at $v_w \approx 0.67$. This peak corresponds to the maximum of 
$\kappa_v \approx 0.44$ in the left panel of Fig.~\ref{fig:kappa}, represented by case B in previous 
discussions, which has supersonic deflagration profile of the plasma surrounding the bubble.
It is clear from this figure
that for a wide range of $v_w$, the SNR for the LISA configuration C1, BBO and UDECIGO is above the two
thresholds $\text{SNR}_{\text{thr}} = 10, 50$. 
For DECIGO, there is also a range 
$0.5 \lesssim v_w < 0.8$ above the threshold 50 and this range becomes much wider for the threshold 10. 
For the LISA configuration C2 with six links, the GW for a wide range 
$0.4 < v_w < 1.0$ is above the threshold value 10 and can therefore be detected according to
Ref.~\cite{Caprini:2015zlo}. 
For the LISA configurations C3 and C4, both of which have four links, 
the uncorrelated noise reduction technique used in the six-link cases is not available and therefore
the SNR needs to be larger than 50 to be detectable~\cite{Caprini:2015zlo}. 
So in this case, the GW is not 
reachable by C3 and C4 for any $v_w$. For ALIA, there is a window at $v_w \approx 0.7$ where the SNR is 
above 10.

\section{Discussion}\label{sec:conclusion}
The discovery of GW at the LIGO initiates a new era in high energy physics and gravity. 
In this paper we propose the stochastic GW as an indirect way of probing the spontaneous breaking new gauge symmetry beyond the SM. 
Working in models with gauged $\mathbf{B-L}$ extension of the SM, we studied the strength of PT relating to the spontaneous breaking of the $\mathbf{B-L}$ as well as the stochastic GW signals generated during the same PT in the space based interferometer.
We find that the power spectrum of GW generated is reachable by the LISA , BBO, ALIA, DECIGO and Ultimate-DECIGO  for the case where the spontaneous breaking of $\mathbf{B-L}$ is triggered by at least two electroweak scalar singlets. 
It should be mentioned that there is no way to identify its intrinsic physics if any stochastic GW signal is observed. 
But it provides a guidance for new physics hunters since stochastic GW signal with peak frequency at near $0.01~{\rm Hz}$  is a hint of new  scalar interactions or new symmetry at the TeV scale.
This work make sense on this point of view. 
Although we only focused on the $U(1)$ case in this paper, our studies can be easily extended to the non-Abelian case since it contains all ingredients for the GW calculation. 

\section*{Acknowledgements}
JS is supported by the National Natural Science Foundation of China under grant 
No.11647601, No.11690022 and No.11675243 
and also supported by the Strategic Priority Research Program of the Chinese Academy of Sciences under 
grant No.XDB23030100. 

\bibliographystyle{utphys}

\end{document}